\documentclass[10pt,sigconf,letterpaper]{acmart}

\usepackage{xspace}
\usepackage{enumitem}
\usepackage{footmisc}
\usepackage{subcaption}
\usepackage{graphicx}
\usepackage{amsmath}
\usepackage{soul}
\usepackage{makecell}

\copyrightyear{2025}
\acmYear{2025}
\setcopyright{cc}
\setcctype{by}
\acmConference[ANRW '25]{Applied Networking Research Workshop}{July 22, 2025}{Madrid, Spain}
\acmBooktitle{Applied Networking Research Workshop (ANRW '25), July 22, 2025, Madrid, Spain}\acmDOI{10.1145/3744200.3744783}
\acmISBN{979-8-4007-2009-3/2025/07}

\begin{document}
\pagestyle{empty}

\title{Locating and Enumerating Anycast: a Comparison of Two Approaches}

 \newif\ifisanon

\iftrue
\newcommand{\raf}[1]{\textcolor{violet}{\noindent[Raf: #1]}}
\else
\newcommand{\raf}[1]{}
\fi
 
\newcommand{\ie}{\textit{i.e.}}
\newcommand{\etc}{\textit{etc.}}
\newcommand{\eg}{\textit{e.g.}}
 
\isanonfalse

\ifisanon

\author{Paper \#408, 13 pages body, 17 pages total} 
\else 

\author{Remi Hendriks}
\affiliation{%
  \institution{University of Twente}
  \city{Enschede}
  \country{the Netherlands}
}
\email{remi.hendriks@utwente.nl}

\author{Tim Betzer}
\affiliation{%
  \institution{Technical University of Munich}
  \city{Munich}
  \country{Germany}
}
\email{betzer@net.in.tum.de}

\author{Ben Du}
\affiliation{%
  \institution{CAIDA/UC San Diego}
  \city{La Jolla, CA}
  \country{USA}
}
\email{bendu@ucsd.edu}

\author{Raffaele Sommese}
\affiliation{%
  \institution{University of Twente}
  \city{Enschede}
  \country{the Netherlands}
}
\email{r.sommese@utwente.nl}

\author{Mattijs Jonker}
\affiliation{%
  \institution{University of Twente}
  \city{Enschede}
  \country{the Netherlands}
}
\email{m.jonker@utwente.nl}

\author{Roland van Rijswijk-Deij}
\affiliation{%
  \institution{University of Twente}
  \city{Enschede}
  \country{the Netherlands}
}
\email{r.m.vanrijswijk@utwente.nl}

\fi

\def \manycasttwo  {MAnycast\textsuperscript{2}\xspace}

\begin{abstract}
    Anycast allows for providing services from multiple, geographically distant
    Points of Presence (PoPs), using a single IP address, to, \eg, improve
    resilience.  Due to its opaqueness, it is often unknown which addresses are
    provisioned using anycast and, if so, where the PoPs are located.  As anycast
    is widely used for critical Internet infrastructures (\eg, the DNS) efforts
    have been made to map anycast deployments.  The current state-of-the-art
    mapping technique, iGreedy, relies on latency-based measurements, and is
    adversely affected by noise caused by, \eg, network processing delays.
    Previous work has shown that traceroute can alternatively be used to detect
    anycast.  As traceroute reveals the hops a packet traverses, it may also be
    used to locate sites using geolocation data for hops near the anycast PoPs. 

    This paper is the first to assess the performance of the traceroute-based approach
    at scale, by targeting 14k prefixes from an anycast census.  Using
    ground truth we show traceroute achieves a slight increase in enumeration and
    geolocation precision over iGreedy.  However, it suffers from overestimating
    the number of PoPs and incurs a 4$\times$ increase in probing cost, making it
    unattractive for anycast censuses.
\end{abstract}



\maketitle

\vspace{-2mm}
\section{Introduction}\label{introduction}
Anycast is the practice of making an IP address available in multiple discrete locations~\cite{rfc4786}.
This allows operators to offer replicated services nearer to clients and increase resilience through redundancy.
Examples include DNS resolvers and nameservers, and CDNs providing web caching.
While the number of anycast addresses on the Internet is small (<0.1\% \cite{manycastr})
they serve a large share of Internet traffic~\cite{anycast_cdn}.

For this reason, efforts have been made to map anycast deployments on the Internet to \eg, measure Internet resilience.
The most notable are iGreedy~\cite{igreedy} and \manycasttwo~\cite{manycast2}, 
where the former, iGreedy, also enumerates and geolocates PoPs behind anycast addresses.
iGreedy geolocates using latency measurements by probing an anycast address from multiple geographically distributed Vantage Points (VPs), similar to conventional IP unicast geolocation~\cite{geolocation_survey}.
However, such latency measurements are known to be noisy due to \eg, network processing and queuing delays.

Unicast IP geolocation often involves the use of tools such as traceroute to improve precision~\cite{geolocation_survey}.
Traceroute allows for measuring the path that an Internet packet takes to reach its target.
When determining the location of a target, %
in case no location information is available for the target itself,
traceroute can be used to find nearby hops for which location information is available.
It has also been used to geolocate anycast using the location of hops near anycast PoPs, to infer the location of PoPs themselves~\cite{regional}.
One example is verifying GDPR compliance when sending personal data to anycast addresses~\cite{hunter}.
However, traceroute has limitations: operators may block the ICMP~Time~Exceeded messages traceroute relies on, and networks that deploy Multiprotocol Label Switching (MPLS) tunneling may hide traceroute hops~\cite{mplscommon}, making it infeasible to geolocate them.

This work investigates the use of traceroute for anycast censuses by performing a large-scale measurement toward 14k anycast prefixes from an anycast census~\cite{manycastr}.
We show traceroute finds an average of 11.82\% more PoPs, yet may also output multiple neighboring locations for a single PoP resulting in overestimation of the number of anycast sites.
Using ground truth we show traceroute achieves a mean geolocation error of 26km compared to 51km using iGreedy.
However, due to a 4$\times$ increase in probing cost, we argue traceroute-based anycast censuses are nonpractical.
We make all code and data publicly available\footnote{\label{fn:git}https://github.com/ut-dacs/anycast-trace-locator}.

This paper is structured as follows.
First, in \S\ref{background} we discuss background on anycast detection and traceroute followed with related work on using traceroute to detect anycast.
Then, we detail the methodology used in \S\ref{methodology} and provide our results in \S\ref{results}.
Finally, we discuss the performance of traceroute to detect anycast and list further use cases in \S\ref{discussion}.


\section{Background and related work}
\label{background}
We present background on traceroute and anycast detection, and discuss work on measuring anycast using traceroute.

\subsection{Traceroute}
Traceroute traces Internet paths by triggering routers on the path towards a target to send back ICMP~\texttt{Time-Exceeded} replies.
Such routers are often configured with DNS \texttt{PTR} records that operators use for debugging purposes,
and may contain information such as the ASN, network type, country and city where the router is located.
IP to location databases may also have known locations for traceroute hops.
This makes it possible to infer (part of) the geographical path of traceroutes. 
Furthermore, it is possible to infer locations of Internet addresses using geolocation information available from nearby hops (\ie, hops with similar RTT values).
In particular, the pen-ultimate hop (\textit{p-hop}), \ie, the hop before the destination, is often used to geolocate IP addresses.

Several works use traceroute to perform unicast geolocation~\cite{geolocation_survey},
in this work we use it to geolocate anycast PoPs.

\subsection{Anycast detection}
The current state-of-the-art technique to detect anycast is iGreedy~\cite{igreedy}.
By measuring the latency from multiple VPs to a target, it finds Great-Circle Distances (GCD) using the speed of packets in fibre optic cables (roughly two thirds the speed of light).
In the case of a unicast target, all circles will overlap providing a single solution that resides in the intersection. 
However, in the case of anycast it finds non-intersecting sets of circles, as VPs reach different anycast PoPs.
In this case, the iGreedy algorithm finds the minimum set of independent overlapping areas in which PoPs must be located for there to be no speed-of-light violation.
Next, it geolocates anycast PoPs 
by selecting an airport near the overlapping area using a metric that includes the population of the main city that the airport serves,
since operators often deploy PoPs in large metropolitan areas to maximize utility.

iGreedy was shown to be quite accurate, achieving a recall of over 50\% with an average geolocation error of 361km.
In the end, the accuracy of iGreedy depends on the number of VPs used and their geographical/topological diversity.

Similarly, RIPE IPmap, a geolocation platform operated by the RIPE NCC, detects anycast using its active geolocation engine.
It classifies an IP address as anycast if multiple globally distributed vantage points observe the target IP address at a latency of 1\,ms or less~\cite{ipmap, ipmapactive}.
Note that this likely leads to a gross underestimation of the number of anycast IPs and sites, since 1\,ms is a very tight bound.

Another detection technique for anycast, \manycasttwo, is achieved by measuring
anycast using anycast\cite{manycast2}.  This lightweight technique does not
allow for geolocating anycast.

\vspace{-2mm}

\subsection{Measuring anycast with traceroute}
Xun et al.~analysed the usage of anycast in the DNS in 2013~\cite{chaos} by
leveraging CHAOS records, which allow for the specific nameserver reached to be
identified~\cite{chaos_rfc}.  As unique CHAOS records were used for multiple
load-balanced nameservers at a single anycast PoP, the authors augmented their
approach with traceroute to resolve ambiguities.

Next, in 2017, Wei et al. measured the occurrence of anycast flipping (\ie, a
single client flipping between multiple anycast PoPs) and used the
pen-ultimate hops of traceroutes to determine the reached PoP~\cite{flipping}.

More recently, in 2023, Zhou et al. used traceroute to geolocate the PoPs of
regional anycast deployments~\cite{regional} By geolocating the p-hop as
observed from RIPE Atlas VPs, they infer the location of the PoP reached by
mapping it to the closest PoP location from available ground truth.  They
performed their methodology towards regional anycast prefixes from two CDNs and
showed they were able to infer the reached site in the majority of cases.

Pascual et al.~introduced a tool to trace anycast communications
(\textit{Hunter}) in 2024, to verify compliance to data protection regulations
for personal data transfers~\cite{hunter}.  Their method geolocates the p-hop
visible in a traceroute, then performs geolocation using latency measurements
from nearby RIPE Atlas VPs.  Their method showed high accuracy and precision in
geolocating the PoP reached, but was only validated using two anycast addresses
from a single operator.


Unlike previous works, which used traceroute towards a select few anycast
deployments, we explore the effectiveness of using traceroute to enumerate and
geolocate anycast PoPs using 14k anycast prefixes from a public anycast census~\cite{manycastr}.


\section{Methodology}\label{methodology}
In this section we describe our methodology that builds on previous work.
Our main contribution is a scalable implementation
that does not rely on \eg, CHAOS records, or ground-truth to cross-reference the PoP reached.

\subsection{Latency propagation}
\begin{figure}[t]
	\centering
	\includegraphics[width=\columnwidth]{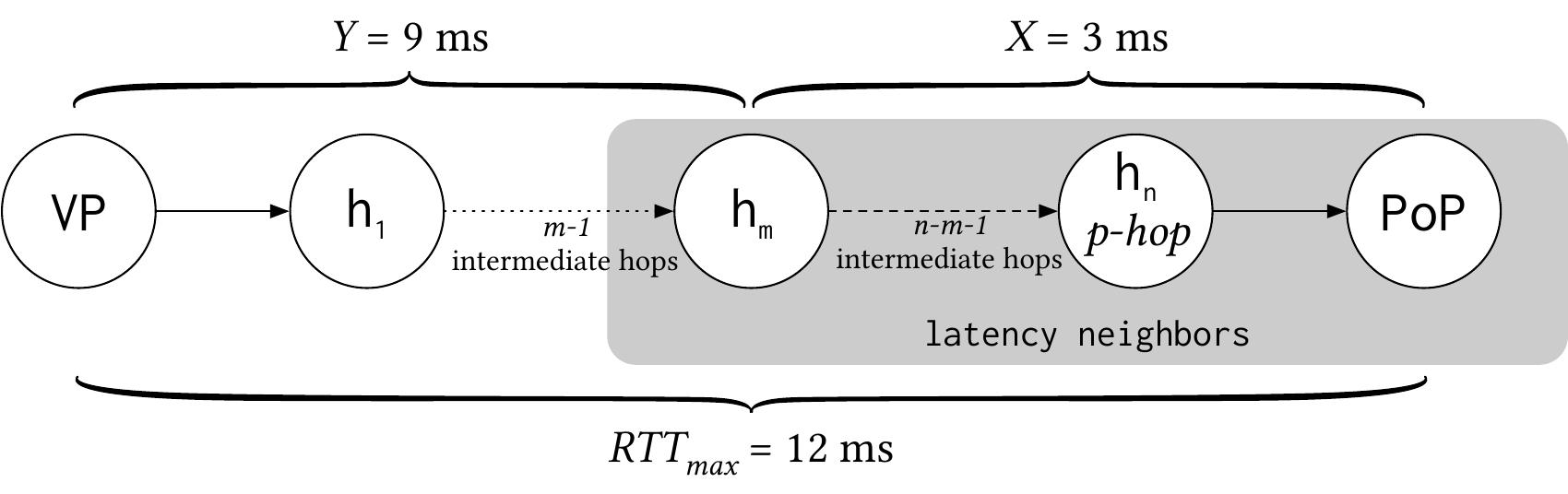}
	\caption{Traceroute latency parameters as per~\cite{traceroutegeolocation}}
	\label{fig:latency-neighbors}
	\vspace{-5mm}
\end{figure}

Latency propagation is a technique used in traditional unicast geolocation,
where traceroute hops with similar round-trip-times (RTTs) are classified as so-called latency neighbors and inferred to be near each other.
Dan et al.~assessed the accuracy in geolocating using latency propagation~\cite{traceroutegeolocation}.
Their methodology uses a latency threshold \textit{X} which determines the maximum RTT difference between two hops for them to be considered latency neighbors.
A second parameter \textit{Y} determines the maximum RTT difference between the VP and the closest latency neighbor.
Their analysis showed that using values of 3\,ms for \textit{X} and 9\,ms for \textit{Y} incurs a median error of 10.1\,km and outperforms IP to location databases.
Using lower parameter values reduces the median error, at a cost in coverage (\ie, for a PoP to be located it requires a VP with less than \textit{X} + \textit{Y} RTT).
We use the same parameters in this work, meaning there must be a latency neighbor within 9\,ms of the VP, with at most 3\,ms RTT difference from the PoP.
This requires the VP to be within 12\,ms of the PoP. Fig.~\ref{fig:latency-neighbors} schematically shows these constraints.

Previous work~\cite{regional, chaos, hunter} uses the penultimate hop \emph{(p-hop)} to infer the location of the PoP.
We later show this approach yields false locations as the \textit{p-hop} may be distant from the PoP (\eg, when there is tunneling). 
Furthermore, the \textit{p-hop} may have no geolocation information available (\eg, it responds with a bogon address)
whereas our approach may use other hops near the PoP (within the latency threshold) that have location data.

\subsection{Locating traceroute hops}
To infer the location of the PoP, we require the location of its latency neighbors.
When the VP itself is a latency neighbor (\ie, within 3ms of the target), we infer the target to be near the VP (which has a known location).
Otherwise, we attempt to geolocate traceroute hops using \texttt{PTR} records.
This is achieved with Hoiho~\cite{hoiho} that extracts geographical information from \texttt{PTR} records using regular expressions obtained using ITDK~\cite{itdk}.
However, \texttt{PTR} records are not always available, may not contain geo-hints, or have an unknown regex pattern in which case we use IPInfo~\cite{ipinfo}, an IP to location database.
We select IPInfo as it shown to be more accurate than other commercial datasets~\cite{ipinfo_best}.
To minimize the impact of wrong geo-hints in \texttt{PTR} records and wrong locations from IP to location databases, we validate hop locations using the measured latency from the VP similar to Zhou et al.~\cite{regional}

\subsection{Inferring PoP location}
Using the latency neighbor technique we may find multiple cities that neighbor a single PoP.
For this reason, we re-use iGreedy's airport selection algorithm~\cite{igreedy} 
that uses the distance from the inferred location and the population of the nearby metropolitan area that the airport serves as heuristics.
We attempt to find an airport within 100\,km from the latency neighbor, if it exists, else we take the nearest airport.

\subsection{Limitations}

\emph{Hidden hops.}
Tunneling and routers configured to not send ICMP~Time~Exceeded replies will result in hidden traceroute hops that may lead to indeterminate results.

\noindent
\emph{Hop latency estimates.}
The RTTs towards traceroute hops are estimates of the actual latency between the VP and the hop itself.
Due to inflated RTTs (caused by \eg, network processing delays), one may infer a hop to be artificially close to the target.
This could lead to false classifications of latency neighbors.
To combat this, we only classify latency neighbors as valid if they are within 9\,ms of the VP as such hops are geographically close to the VP and generally suffer less from network processing delays.

\vspace{0.5em}
\noindent
\emph{Grouping locations.}
When performing a traceroute from multiple VPs to a single target, \eg, a single PoP, we will find multiple neighboring routers surrounding the target.
These routers may be at most 6\,ms distant from each other, as a latency neighbor is within 3\,ms of the target, this roughly translates to a possible distance of 600\,km.
To avoid counting multiple PoPs in such situations, we group results using iGreedy's algorithm that locates the most likely nearby airport based on its distance and population it serves.
However, we may find multiple major airports surrounding a single PoP where our method overestimates the number of PoPs.

To avoid the possibility of overcounting, in the worst case, requires to group locations found within 600\,km of each other as a single PoP.
Yet, by doing so it severely reduces performance as PoPs within such radius are no longer distinguishable (\eg, the distance between Amsterdam and Berlin is less than 600\,km).
We later show that using a radius of 100\,km provides the best trade-off between performance and overestimating PoPs using ground truth.
 
\vspace{-2mm}

\subsection{Measurement setup}
We use all VPs of CAIDA's \textit{Archipelago} (Ark), covering 66 countries, from which we perform traceroute measurements.
We choose this platform as it provides accurate locations of VPs, has good global coverage, and allows for large scale traceroute measurements~\cite{arkdsl}.
Furthermore, Ark implements Paris traceroute~\cite{paris} which avoids anomalous traceroute results due to multi-path topologies caused by load-balancers.
Using Ark we send a single probe per traceroute hop, unlike traditional traceroute that sends 3 probes per hop, to limit the probing burden.
This analysis can be repeated using RIPE Atlas, and other probing platforms, though it would require filtering of Atlas probes with wrong user reported locations.

We recall that our goal is to analyse how well traceroute can enumerate and geolocate PoPs of an anycast deployment.
In order to reduce the impact of our experiments, we therefore chose to target a list of 14k known anycast prefixes that were confirmed using iGreedy~\cite{manycastr}.
Furthermore, we compare against our own iGreedy measurement using GCD from ping latencies, which we run using the same VPs offered by Ark.
Finally, we make use of publicly available information of root servers to determine the precision of both methodologies in geolocating PoPs~\cite{root_locations}.


\section{Results}\label{results}
In total, our dataset consists of traceroutes measured between 274 VPs (inside 178 distinct ASes) and 13,765 target anycast /24-prefixes.
For 30 prefixes no completed traceroutes were captured, \eg, because the AS or one of its upstreams blocks ICMP~Time~Exceeded messages,
a known limitation of traceroute.
Manual inspection shows they are ping responsive but unreachable with traceroute.
The remaining 13,735 targets (inside 906 distinct ASes) consist of 3.35~million traceroutes, averaging 244 completed traceroutes per target.
The total number of traceroute hops captured is 27.2~million, averaging 8.12 hops captured for each \textit{(VP, target)} pair.

In total, our traceroutes traverse 1,827 distinct ASes and we observe 99,939 unique hop addresses (including 6,282 bogon addresses).
Using OpenINTEL~\cite{openintel} we obtain \texttt{PTR} records for 45,106 addresses
of which 19,613 (43.5\%) were translated using Hoiho~\cite{hoiho} and 16,565 (36.7\%) contained city-level geolocation.
Next, using IPInfo we obtain city-level geolocation for all non-bogon addresses.
We validate locations by assessing whether it is within possible distance of the VP using the measured RTT towards that hop.
Whenever a valid location is available we use it, preferring \texttt{PTR} record locations over IPInfo.
In total, we have 16,929 (16.9\%) hops with a valid PTR location and 75,143 (75.2\%) with a valid IPInfo location.
For 7,867 (7.9\%) hops there is no valid location available.

In total, we confirm 13,478 prefixes to be anycast using the latency neighbor methodology.
For the 257 missing prefixes, 
197 are observed to have two PoPs using GCD and 29 to have three PoPs.
These are mostly easy to detect cases of anycast using GCD, as the PoPs are geographically distant from each other.
However, they only have an Ark VP within 12 ms for at most one PoP (\ie, the remaining PoPs are not visible with the latency neighbor method).
The remaining 31 prefixes failed due to a low number of VPs providing completed traceroutes.

\subsection{Enumeration}
\begin{figure}[t]
  \centering
  \includegraphics[width=\columnwidth]{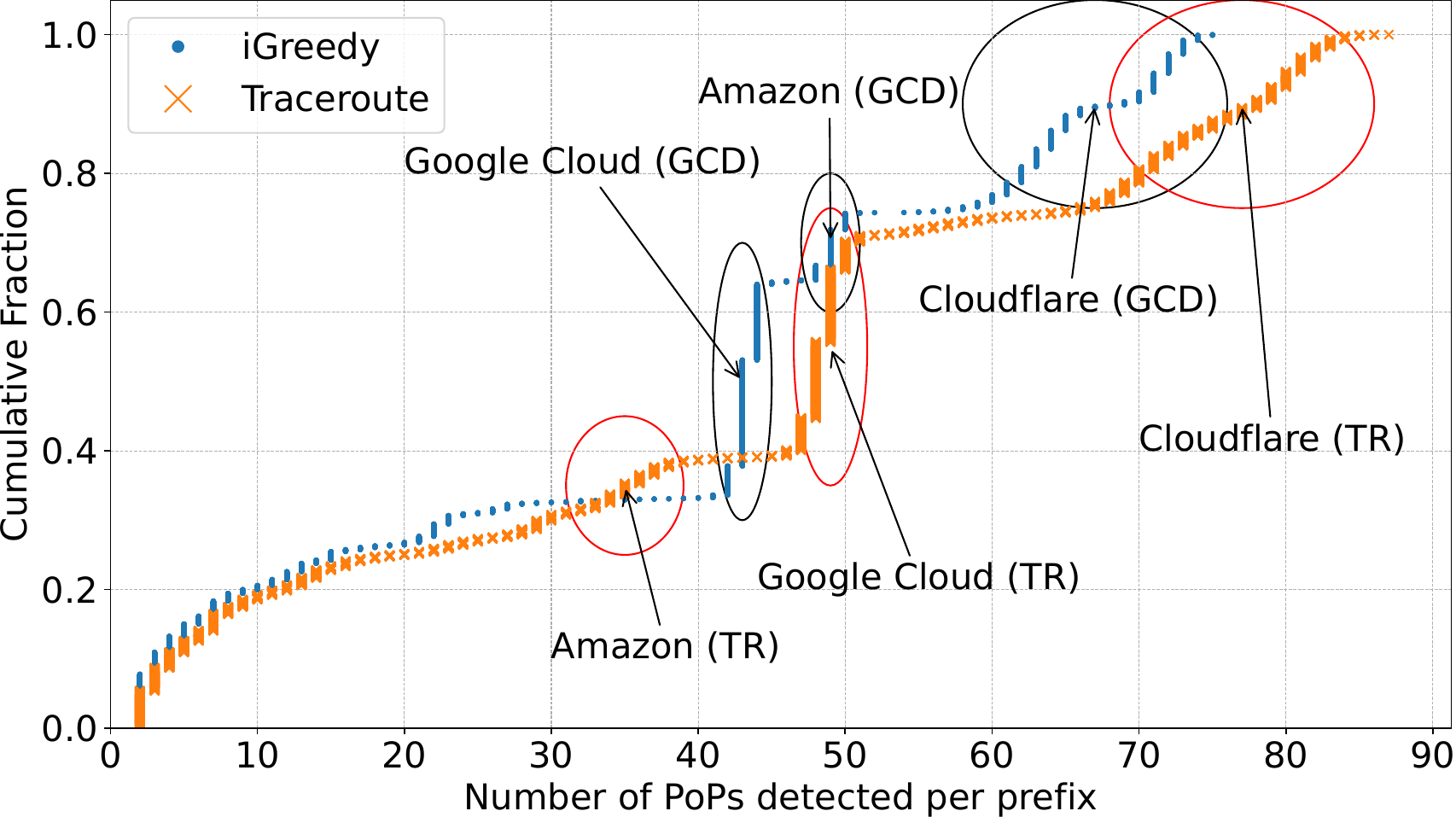}
  \caption{
  Distribution of PoPs found using the iGreedy and traceroute technique for 13,765 anycast prefixes.
}
 \label{fig:enumeration}
 \vspace{-5mm}
\end{figure}

Fig.~\ref{fig:enumeration} shows the CDF for the number of PoPs detected by both methodologies.
We find that for small to medium sized anycast deployments (< 30 PoPs, bottom left) both methodologies detect a similar number of sites.
Next, for large deployments (> 40 PoPs, top right) we observe that traceroute consistently finds more PoPs.
Overall, the traceroute technique finds 11.82\% more PoPs compared to iGreedy.

Most of the large deployments can be attributed to a few CDNs.
Since the additional benefit of traceroute in terms of enumeration is most visible for large deployments,
Table~\ref{tab:hypergiants} provides a break-down on the enumeration counts for the 5 largest ASes in our dataset alongside ground truth.
For ground truth we use public information on the number of cities where PoPs are deployed.
Operators may deploy prefixes using different configurations including different numbers of PoPs used.
For example, Fastly has PoPs in 79 distinct cities~\cite{fastly_locations} but offers \textit{intelligent PoP placement}~\cite{fastly_placement} where selective deployments are offered to customers.

We find the traceroute technique achieves a higher enumeration count for all CDNs except Amazon.
Investigating these targets, we observe no traceroute hops are visible inside Amazon's network.
As our methodology requires a latency neighbor within 3\,ms of the target, this means that any routing longer than 3\,ms inside Amazon's network results in an undetected PoP.
 
\begin{table}[t]
\footnotesize
\resizebox{\columnwidth}{!}{
\begin{tabular}{|c|c|r|r|r|r|}
    \hline
     \textbf{AS} & \textbf{Organization} & \textbf{IPv4 (/24)} & \textbf{TR} & \textbf{GCD} & \textbf{GT} \\
    \hline
    \textbf{396982} & Google Cloud & 3,915 & 48.4 & 43.2 & 103~\cite{google_locations} \\
    \hline
    \textbf{13335} & Cloudflare & 3,096 & 74.2 & 65.9 & 335~\cite{cloudflare_locations} \\
    \hline
    \textbf{16509} & Amazon & 1,325 & 40.2 & 48.6 & >300~\cite{amazon_locations} \\
    \hline
    \textbf{54113} & Fastly &  799 & 16.5 & 13.4 & 79~\cite{fastly_locations} \\
    \hline
    \textbf{209242} & Cloudflare Spectrum & 306 & 74.6 & 63.7 & 335~\cite{cloudflare_locations} \\
    \hline
\end{tabular}=
} 
\vspace{0.1em}
\caption{
	Largest ASes originating anycast with the mean PoP count found using traceroute (TR), iGreedy (GCD), and the PoP count from public data (GT).
}
\vspace{-2em}
\label{tab:hypergiants}
\end{table} 

\subsection{Geolocation}

\begin{figure}[t]
  \centering
  \includegraphics[width=0.8\columnwidth]{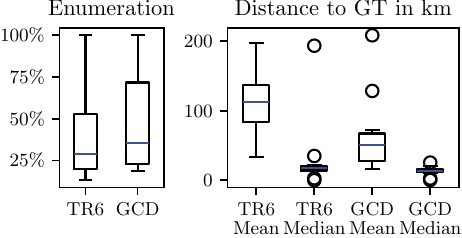}
  \vspace{5pt}\\
  \includegraphics[width=0.8\columnwidth]{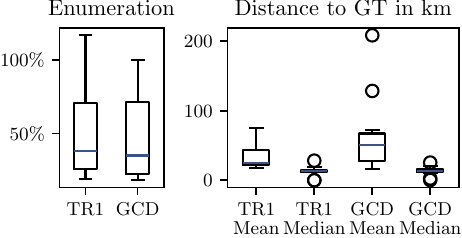}
  \caption{
    Comparison between traceroute with 600\,km (TR6) and 100\,km (TR1) radius and iGreedy (GCD) results using the root server locations as ground truth (GT). Enumeration compares identified locations to GT; values over 100\% indicate overcounting.
  }
 \label{fig:groundtruth}
 \vspace{-2mm}
\end{figure}

Fig.~\ref{fig:groundtruth} compares traceroute to iGreedy using all DNS root servers as ground truth except the G-root server since it is unresponsive to pings. The left-side of the figure shows a boxplot for enumeration counts, and we show the mean and median errors for geolocation on the right.
As discussed previously we compare results when grouping using a 600\,km radius (top) and a 100\,km radius (bottom)

For iGreedy we find a recall of 35.19\% PoPs detected and no overcounting.
The median for mean distances is 50.81\,km and for median distances it is 12.94\,km.
Traceroute with a grouping radius of 600\,km has a median enumeration of 28.97\% with no overcounting.
The median for mean distances is more than double compared to iGreedy with 112.18\,km and for median distances amounts to 17.02\,km.
Next, a grouping radius of 100\,km leads to overcounting in two instances 
with a median enumeration of 38.53\%,
while reducing the median for mean distances to 25.58\,km and to 13.57\,km for median distances.
Overall, we find that the grouping radius provides a trade-off where a lower radius yields more accurate geolocation results,
but increases the occurrence of overcounting.
Despite the overcounting, we use a 100\,km grouping factor as it yields more accurate results.
We provide a breakdown of results per root-letter in Table~\ref{tab:roots}.

\begin{table}[t]
\centering
\footnotesize
\begin{tabular}{lrrrrr}
\toprule
\textbf{Letter} & \textbf{GT} & \textbf{\makecell{GCD\\Count}} & \textbf{\makecell{TR\\Count}} & \textbf{\makecell{GCD\\Geo error}} & \textbf{\makecell{TR\\Geo error}} \\
\midrule
A & 20 & 0.7000 & 0.7500 & 15.7548 & 42.7639 \\
B & 6 & 1.0000 & \textcolor{red}{1.1667} & 22.6511 & 54.0090 \\
C & 13 & 0.7692 & 0.6923 & 59.6628 & 47.5731 \\
D & 222 & 0.2748 & 0.3108 & 18.2553 & 22.7183 \\
E & 325 & 0.2308 & 0.2615 & 29.8686 & 21.0821 \\
F & 359 & 0.1838 & 0.2089 & 61.9196 & 21.7074 \\
H & 12 & 1.0000 & \textcolor{red}{1.0833} & 208.6422 & 75.8460 \\
G & 6 & 0.0000 & 0.0000 & - & - \\
I & 85 & 0.3176 & 0.3647 & 72.0640 & 29.2808 \\
J & 101 & 0.3861 & 0.4059 & 41.9601 & 25.2775 \\
K & 121 & 0.2231 & 0.1901 & 128.5593 & 19.2221 \\
L & 123 & 0.2033 & 0.2602 & 34.7577 & 23.8902 \\
M & 23 & 0.6087 & 0.5217 & 65.7767 & 17.2038 \\
\bottomrule
\end{tabular}
\vspace{0.1em}
\caption{
	Breakdown of results per Root Server (Letter) with number of PoPs (GT), ratio of PoPs found (GCD, TR Count) and mean geographical error (GCD, TR Geo). Red indicates overcounting.
}
\vspace{-2em}
\label{tab:roots}
\vspace{-2mm}
\end{table}

\subsection{Probing cost}
A large limitation of traceroute is probing cost as traceroute requires multiple packets to measure a path.
In our dataset we observe an average of 9.83 hops including the destination.
Compared to iGreedy, which measures the latency with a single ping packet this is nearly a 10-fold increase in probing cost.
However, we can reduce the probing cost by skipping initial hops adjacent to the VP and inside the same AS. 
On average, we find an average of 2.87 of same-AS hops lowering the mean count to 6.96 hops.
This is a maximum on the number of hops that may be skipped, as VPs may have an inconsistent number of same-AS hops. 

Next, we can reduce the probing cost even further as we require a measured RTT below 12\,ms between the VP and target (to satisfy the latency neighbor constraints).
Therefore, using an initial ping measurement from all VPs we can limit traceroutes to targets within this threshold.
On average we find 124 out of 274 VPs are within 12\,ms of the anycast target.
This constraint also reduces the average hop count decreases from 9.83 to 8.40, as these VPs are closer to the target,
which can be reduced to 5.53 when skipping same-AS hops.

We provide a breakdown of the probing cost in Table~\ref{tab:cost}.
This shows the probing cost can be reduced to 795 probes per target, by performing a single latency measurement from all 274 probes,
followed with a traceroute measurement from an average of 124 VPs within 12\,ms that measure an average of 4.20 hops when skipping same-AS hops.
Unsurprisingly, we find that VPs with a low latency traverse a lower number of hops to reach the target.

\begin{table}[t]
\centering
\footnotesize
\begin{tabular}{|l|r|r|r|}
    \hline
     \textbf{Technique} & \textbf{Probes} & \textbf{\# of VPs} & \textbf{Total} \\
    \hline
    \textbf{TR (full)} & 9.83{\textcolor{white}*} & 274\textcolor{white}{$^\dagger$} & 2,693 \\
    \hline
    \textbf{TR (skipping same-AS hops)} & 6.96{\textcolor{white}*} & 274\textcolor{white}{$^\dagger$} & 1,907 \\
    \hline
    \textbf{TR (latency constrained)} & 8.40* & 124$^\dagger$ & 1,316 \\ 
    \hline
    \textbf{TR (reduced)} & 5.53* & 124$^\dagger$ & 960 \\ 
    \hline
    \hline
    \textbf{iGreedy} & 1{\textcolor{white}*} & 274\textcolor{white}{$^\dagger$} & 274 \\
    \hline
    \multicolumn{4}{l}{*Initial measurement required from all 274 VPs.} \\
    \multicolumn{4}{l}{$^\dagger$Average number of VPs within 12\,ms RTT.}
\end{tabular}
\vspace{0.1em}
\caption{
	Probing cost per target breakdown.
}
\vspace{-2em}
\label{tab:cost}
\vspace{-5mm}
\end{table} 



\subsection{Using p-hops}\label{mapping}
As mentioned, previous work used the \textit{p-hop} to assess the location of the anycast site reached.
However, we only use \textit{latency neighbors} that are within 3\,ms of the target for VPs with at most 12\,ms RTT towards the target.
To assess the accuracy of relaxing these latency constraints to use all \textit{p-hops} measured, we perform an unconstrained measurement towards the DNS root letters.
On average, this finds 11.92 more sites for each root letter. 
Yet it also amplifies the overestimation error motivating our decision to use the latency neighbor constraints.
For example, B-root has 6 PoPs where the unconstrained \textit{p-hop} method finds 13 PoPs and for H-root with 12 PoPs it finds 30. 

\begin{figure}[t]
  \centering
  \includegraphics[width=0.9\columnwidth]{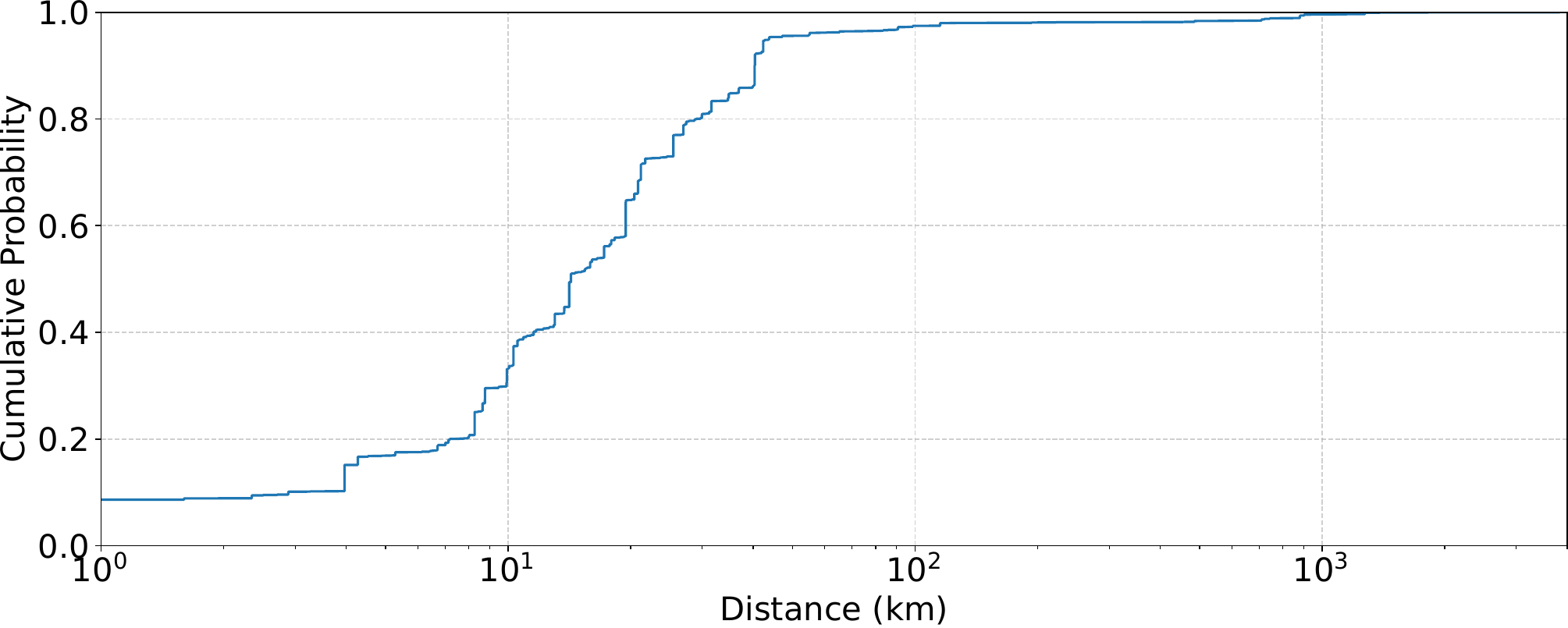}
  \caption{
  Distances (log scale) between invalid latency neighbor \textit{p-hops} and the nearest airport found using the latency neighbor method.
}
 \label{fig:phop_dis}
 \vspace{-5mm}
\end{figure}

However, similar to Zhou et al.~we can cross-reference the PoP reached using \textit{p-hop} locations~\cite{regional}.
In total, we have 3.35 million \textit{p-hops} (averaging 243 per target) in our dataset, 1.88 million of which are invalid latency neighbors.
We cross-reference these invalid hops with inferred locations using valid latency neighbors.
Figure~\ref{fig:phop_dis} shows the average distance between invalid \textit{p-hops} and inferred locations.
Overall, we find that the majority of \textit{p-hops} can be correlated to a PoP found within a 40\,km radius,
therefore this method is valid when cross-referencing locations with \eg, ground-truth or iGreedy locations.
However, a long tail with large distances cannot be correlated to a nearby PoP location
that may lead to overcounting when used to infer additional locations.


\vspace{-3mm}

\section{Discussion}\label{discussion}
Our findings show that traceroute outperforms iGreedy in terms of enumeration and geolocation.
More specifically, we find that traceroute achieves an average increase of 11.82\% in enumerating anycast PoPs and has a lower mean geolocation error (26\,km compared to 51\,km).
However, due to, \eg, unreliable hop latencies it may overestimate the number of PoPs and it comes at a significant increase in probing cost compared to iGreedy.
Therefore, we do not recommend to use traceroute as a method to perform anycast censuses.

Our results do indicate other benefits to performing traceroutes to anycast, such as mapping which ASes reach a particular PoP
and mapping topological properties of anycast deployments (\eg, the upstream ASes for a particular PoP).

\subsection{Future work}

\paragraph{Longitudinal analysis}
Like IPMap~\cite{ipmap}, our methodology can be performed using historical traceroute data.
RIPE Atlas, Measurement Lab (M-Lab), and Ark have public longitudinal datasets of traceroute measurements.
Most notable is FANTAIL~\cite{fantail} which provides access to historical traceroute data towards all routable /24 IPv4 prefixes~\cite{traceroute_campaign}.
Our analysis can be repeated using such traceroute data to map the development of anycast infrastructures over time, giving insights in, \eg, infrastructure expansions and topological changes.

\paragraph{Mapping anycast topologies}
We can infer the ASes traversed in traceroutes and identify upstream ASes for particular PoPs using CAIDA's prefix2as dataset~\cite{prefix2as}.
Using this method we can find a lower-bound on the upstream ASes present at each anycast PoP.
With \eg, RIPE Atlas, this analysis can be performed from a large number of origin ASes to obtain a more complete picture.

\paragraph{Sub-optimal anycast routing}
As done by Rizvi et al.~\cite{polarization} traceroute can be used to detect causes of sub-optimal anycast routing (\ie, a client routing to a distant anycast site when a nearby one is available).
Our methodology can extend their work as operators can use it to infer the PoP receiving traffic from distant clients.
Furthermore, whilst Rizvi et al.~looked at the penultimate AS, we observe adjacent hops in distinct ASes with large latency increases on-path to the anycast AS.
We suspect these are remote-peering links.

\paragraph{Further refinement}
The average distance between p-hops (that are invalid latency neighbors) and the inferred anycast location suggest that the latency neighbor constraints may be too restrictive.
Furthermore, false latency neighbors can possibly be avoided by performing multiple traceroutes and taking the minimum of observed latencies to each hop.
However, this would come at a significant increase in probing cost.
Finally, the grouping radius may be altered to improve geolocation accuracy at the cost of overcounting, or vice versa.
We leave these parameters adjustable in our code\footnote{\label{fn:git}https://github.com/ut-dacs/anycast-trace-locator}.

\begin{acks}
The CAIDA Ark platform is supported by U.S. NSF grants OAC-2131987, CNS-2120399, and CNS-2212241. 
The views and conclusions are those of the authors and do not necessarily represent endorsements, 
either expressed or implied, of NSF.
This work has been partially funded by CATRIN (NWO grant NWA.1215.18.003).
This research was made possible by OpenINTEL, a joint project of the University of Twente, SURF, SIDN, and NLnet Labs.
\end{acks}

\bibliographystyle{ACM-Reference-Format}
\bibliography{atr}

@article{hunter,
title = {Hunter: Tracing anycast communications to uncover cross-border personal data transfers},
journal = {Computers \& Security},
volume = {141},
pages = {103823},
year = {2024},
issn = {0167-4048},
doi = {https://doi.org/10.1016/j.cose.2024.103823},
url = {https://www.sciencedirect.com/science/article/pii/S016740482400124X},
author = {Hugo Pascual and Jose M. {del Alamo} and David Rodriguez and Juan C. Dueñas}
}

@inproceedings{regional,
author = {Zhou, Minyuan and Zhang, Xiao and Hao, Shuai and Yang, Xiaowei and Zheng, Jiaqi and Chen, Guihai and Dou, Wanchun},
title = {Regional IP Anycast: Deployments, Performance, and Potentials},
year = {2023},
isbn = {9798400702365},
publisher = {Association for Computing Machinery},
url = {https://doi.org/10.1145/3603269.3604846},
doi = {10.1145/3603269.3604846},
booktitle = {Proceedings of the ACM SIGCOMM Conference},
pages = {917–931},
numpages = {15},
series = {ACM SIGCOMM '23}
}

@misc{ipmap,
	author = {RIPE NCC},
	title = {IPmap},
	howpublished = {\url{https://ipmap.ripe.net/}},
	year = {2025},
	note = {[Accessed 29-04-2025]},
}

@INPROCEEDINGS{chaos,
  author={Fan, Xun and Heidemann, John and Govindan, Ramesh},
  booktitle={2013 Proceedings IEEE INFOCOM}, 
  title={Evaluating anycast in the domain name system}, 
  year={2013},
  volume={},
  number={},
  pages={1681-1689},
  doi={10.1109/INFCOM.2013.6566965}}

@ARTICLE{flipping,
  author={Wei, Lan and Heidemann, John},
  journal={IEEE Transactions on Network and Service Management}, 
  title={Does Anycast Hang Up on You (UDP and TCP)?}, 
  year={2018},
  volume={15},
  number={2},
  pages={707-717},
  doi={10.1109/TNSM.2018.2804884}}

@misc{rfc4786,
    series =    {Request for Comments},
    number =    4786,
    howpublished =  {RFC 4786},
    publisher = {RFC Editor},
    doi =       {10.17487/RFC4786},
    url =       {https://rfc-editor.org/info/rfc4786},
    author =    {Kurt Erik Lindqvist and Joe Abley},
    title =     {{Operation of Anycast Services}},
    pagetotal = 24,
    year =      2006,
    month =     dec,
}

@inproceedings{manycast2,
author = {Sommese, Raffaele and Bertholdo, Leandro and Akiwate, Gautam and Jonker, Mattijs and van Rijswijk-Deij, Roland and Dainotti, Alberto and Claffy, KC and Sperotto, Anna},
title = {MAnycast2: Using Anycast to Measure Anycast},
year = {2020},
isbn = {9781450381383},
publisher = {Association for Computing Machinery},
url = {https://doi.org/10.1145/3419394.3423646},
doi = {10.1145/3419394.3423646},
booktitle = {Proceedings of the 20th ACM IMC},
pages = {456–463},
numpages = {8},
series = {IMC '20}
}

@INPROCEEDINGS{igreedy,
  author={Cicalese, Danilo and Joumblatt, Diana and Rossi, Dario and Buob, Marc-Olivier and Augé, Jordan and Friedman, Timur},
  booktitle={2015 IEEE Conference on Computer Communications (INFOCOM)}, 
  title={A fistful of pings: Accurate and lightweight anycast enumeration and geolocation}, 
  year={2015},
  volume={},
  number={},
  pages={2776-2784},
  doi={10.1109/INFOCOM.2015.7218670}}

@misc{anycast_cdn,
      title={A First Look at Anycast CDN Traffic}, 
      author={Danilo Cicalese and Danilo Giordano and Alessandro Finamore and Marco Mellia and Maurizio Munafò and Dario Rossi and Diana Joumblatt},
      year={2021},
      eprint={1505.00946},
      archivePrefix={arXiv},
      primaryClass={cs.NI},
      url={https://arxiv.org/abs/1505.00946}, 
}

@inproceedings{traceroutegeolocation,
author = {Dan, Ovidiu and Parikh, Vaibhav and Davison, Brian D.},
title = {IP Geolocation Using Traceroute Location Propagation and IP Range Location Interpolation},
year = {2021},
isbn = {9781450383134},
publisher = {Association for Computing Machinery},
url = {https://doi.org/10.1145/3442442.3451888},
doi = {10.1145/3442442.3451888},
booktitle = {Companion Proceedings of the Web Conference},
pages = {332–338},
numpages = {7},
series = {WWW '21}
}

@inproceedings{hoiho,
author = {Luckie, Matthew and Huffaker, Bradley and Marder, Alexander and Bischof, Zachary and Fletcher, Marianne and Claffy, K},
title = {Learning to extract geographic information from internet router hostnames},
year = {2021},
isbn = {9781450390989},
publisher = {Association for Computing Machinery},
url = {https://doi.org/10.1145/3485983.3494869},
doi = {10.1145/3485983.3494869},
booktitle = {Proceedings of the 17th International CoNEXT},
pages = {440–453},
numpages = {14},
series = {CoNEXT '21}
}

@misc{ipinfo,
	author = {ipinfo.io},
	title = {Trusted IP Data Provider, from IPv6 to IPv4},
	howpublished = {\url{https://ipinfo.io}},
	note = {[Accessed 29-04-2025]},
	year = {2025}
}

@inproceedings{paris,
author = {Augustin, Brice and Cuvellier, Xavier and Orgogozo, Benjamin and Viger, Fabien and Friedman, Timur and Latapy, Matthieu and Magnien, Cl\'{e}mence and Teixeira, Renata},
title = {{Avoiding Traceroute Anomalies with Paris Traceroute}},
year = {2006},
isbn = {1595935614},
booktitle = {Proceedings of the 6th ACM IMC},
pages = {153–158},
numpages = {6},
}

@techreport{chaos_rfc,
  author = "Woolf, S. and D. Conrad",
  title = "{Requirements for a Mechanism Identifying a Name Server Instance}",
  type = "RFC",
  number = "4892",
  year = "2007",
  url = "https://rfc-editor.org/rfc/rfc4892",
}

@misc{prefix2as,
	author = {CAIDA},
	title = {{R}outeviews {P}refix to {A}{S} mappings {D}ataset (pfx2as) for {I}{P}v4 and {I}{P}v6 --- caida.org},
	howpublished = {\url{https://caida.org/catalog/datasets/routeviews-prefix2as}},
	year = {2024},
	note = {[Accessed 29-04-2025]},
}

@misc{google_locations,
	author = {Google},
	title = {Network edge locations},
	howpublished = {\url{https://cloud.google.com/vpc/docs/edge-locations}},
	year = {2025},
	note = {[Accessed 25-04-2025]},
}

@misc{cloudflare_locations,
	author = {Cloudflare},
	title = {Network edge locations},
	howpublished = {\url{https://cloudflare.com/network}},
	year = {2025},
	note = {[Accessed 29-04-2025]},
}

@misc{amazon_locations,
	author = {Amazon (AWS)},
	title = {Global Edge Network},
	howpublished = {\url{https://aws.amazon.com/cloudfront/features}},
	year = {2025},
	note = {[Accessed 25-04-2025]},
}

@misc{fastly_locations,
	author = {Fastly},
	title = {Network Map},
	howpublished = {\url{https://www.fastly.com/network-map}},
	year = {2025},
	note = {[Accessed 25-04-2025]},
}

@misc{fastly_placement,
	author = {Fastly},
	title = {Managed CDN},
	howpublished = {\url{https://www.fastly.com/services/managed-cdn}},
	year = {2025},
	note = {[Accessed 25-04-2025]},
}

@misc{root_locations,
	author = {root-servers.org},
	title = {Root server locations},
	howpublished = {\url{https://root-servers.org/}},
	year = {2025},
	note = {[Accessed 25-04-2025]},
}

@misc{manycastr,
      title={MAnycast Reloaded: a Tool for an Open, Fast, Responsible and Efficient Daily Anycast Census}, 
      author={Remi Hendriks and Matthew Luckie and Mattijs Jonker and Raffaele Sommese and Roland van Rijswijk-Deij},
      year={2025},
      eprint={2503.20554},
      archivePrefix={arXiv},
      primaryClass={cs.NI},
      url={https://arxiv.org/abs/2503.20554}, 
}

@article{mplscommon,
author = {Donnet, Benoit and Luckie, Matthew and M\'{e}rindol, Pascal and Pansiot, Jean-Jacques},
title = {Revealing MPLS tunnels obscured from traceroute},
year = {2012},
issue_date = {April 2012},
publisher = {Association for Computing Machinery},
volume = {42},
number = {2},
issn = {0146-4833},
url = {https://doi.org/10.1145/2185376.2185388},
doi = {10.1145/2185376.2185388},
journal = {SIGCOMM Comput. Commun. Rev.},
month = mar,
pages = {87–93},
numpages = {7}
}

@article{ipmapactive,
author = {Du, Ben and Candela, Massimo and Huffaker, Bradley and Snoeren, Alex C. and claffy, kc},
title = {RIPE IPmap active geolocation: mechanism and performance evaluation},
year = {2020},
issue_date = {April 2020},
publisher = {Association for Computing Machinery},
volume = {50},
number = {2},
issn = {0146-4833},
url = {https://doi.org/10.1145/3402413.3402415},
doi = {10.1145/3402413.3402415},
journal = {SIGCOMM Comput. Commun. Rev.},
month = may,
pages = {3–10},
numpages = {8}
}

@ARTICLE{geolocation_survey,
  author={Zilberman, Aviram and Offer, Adi and Pincu, Bar and Glickshtein, Yoni and Kant, Roi and Brodt, Oleg and Otung, Andikan and Puzis, Rami and Shabtai, Asaf and Elovici, Yuval},
  journal={IEEE Communications Surveys \& Tutorials}, 
  title={A Survey on Geolocation on the Internet}, 
  year={2024},
  volume={},
  number={},
  pages={1-1},
  doi={10.1109/COMST.2024.3518398}}

@misc{fantail,
  title = {{FANTAIL: Facilitating Advances in Network Topology Analysis}},
  venue = {Workshop on Active Internet Measurements: Knowledge of Internet Structure: Measurement, Epistemology, and Technology (AIMS-KISMET)},
  month = {February},
  year = {2020},
  howpublished = {\url{https://catalog.caida.org/presentation/2020_fantail_kismet}},
  note = {Accessed: 29-04-2025},
  doi = {https://catalog.caida.org/presentation/2020_fantail_kismet}
}

@INPROCEEDINGS{traceroute_campaign,
  author={Claffy, Kimberly and Hyun, Young and Keys, Ken and Fomenkov, Marina and Krioukov, Dmitri},
  booktitle={2009 Cybersecurity Applications \& Technology Conference for Homeland Security}, 
  title={Internet Mapping: From Art to Science}, 
  year={2009},
  volume={},
  number={},
  pages={205-211},
  doi={10.1109/CATCH.2009.38}}

@inproceedings{ipinfo_best,
author = {Darwich, Omar and Rimlinger, Hugo and Dreyfus, Milo and Gouel, Matthieu and Vermeulen, Kevin},
title = {Replication: Towards a Publicly Available Internet Scale IP Geolocation Dataset},
year = {2023},
isbn = {9798400703829},
publisher = {Association for Computing Machinery},
url = {https://doi.org/10.1145/3618257.3624801},
doi = {10.1145/3618257.3624801},
booktitle = {Proceedings of the 23rd ACM IMC},
pages = {1–15},
numpages = {15},
series = {IMC '23}
}

@InProceedings{polarization,
author="Rizvi, A. S. M.
and Huang, Tingshan
and Esrefoglu, Rasit
and Heidemann, John",
editor="Richter, Philipp
and Bajpai, Vaibhav
and Carisimo, Esteban",
title="Anycast Polarization in the Wild",
booktitle="Passive and Active Measurement",
year="2024",
publisher="Springer Nature Switzerland",
pages="104--131",
isbn="978-3-031-56252-5"
}

@inproceedings{arkdsl,
  title={An Integrated Active Measurement Programming Environment},
  author={Luckie, Matthew and Hariprasad, Shivani and Sommese, Raffaele and Jones, Brendon and Keys, Ken and Mok, Ricky and Claffy, K},
  booktitle={International Conference on Passive and Active Network Measurement},
  pages={137--152},
  year={2025},
  organization={Springer}
}

@ARTICLE{openintel,
  author={van Rijswijk-Deij, Roland and Jonker, Mattijs and Sperotto, Anna and Pras, Aiko},
  journal={IEEE JSAC}, 
  title={A High-Performance, Scalable Infrastructure for Large-Scale Active DNS Measurements}, 
  year={2016},
  volume={34},
  number={6},
  pages={1877-1888},
  doi={10.1109/JSAC.2016.2558918}
  }

@misc{itdk,
	author = {CAIDA},
	title = {Macroscopic Internet Topology Data Kit --- caida.org},
	howpublished = {\url{https://www.caida.org/catalog/datasets/internet-topology-data-kit/release-2024-08/}},
	year = {2024},
	note = {[Accessed 13-06-2025]},
}


\end{document}